\documentclass[aps,prl,superscriptaddress,twocolumn]{revtex4-1}

\usepackage{color}

\usepackage{amsmath,amsfonts,amsthm}
\usepackage{chngcntr}
\usepackage{gensymb}
\usepackage{amssymb}

\usepackage{soul}
\usepackage[T1]{fontenc} 
\usepackage[english]{babel} 
\usepackage{multirow}
\usepackage{graphicx} 

\usepackage{lmodern}
\usepackage{mathtools}
\usepackage{hyperref}   

\usepackage{lipsum} 
\usepackage{fancyhdr} 
\numberwithin{equation}{section} 
\numberwithin{figure}{section} 
\numberwithin{table}{section} 

\counterwithout{figure}{section}
\counterwithout{table}{section}
\begin{document}
\newcommand{\hbcomm}[1]{\textcolor{blue}{\textit{\textbf{HB:} #1}}}
\newcommand{\hb}[1]{\textcolor{blue}{#1}}
\title{Ligand hole driven metal-insulator transition in a prototypical transition metal double perovskite oxide Ca$_2$FeMnO$_6$ } 


\author{ Arindam Sarkar}
\affiliation{Department of Physics, Indian Institute of Technology, Bombay, Powai, Mumbai 400076, India} 
\author{Hrishit Banerjee}
\affiliation{School of Science and Engineering, University of Dundee, Nethergate, Dundee, DD1 4HN, Scotland, UK} 
\affiliation{Yusuf Hamied Department of Chemistry, University of Cambridge, Cambridge CB2 1EW, UK} 
\author{Debashish Das}
\affiliation{School of Basic Sciences, Indian Institute of Technology Bhubaneswar, Khordha 752050 Odisha, India} 
\author{Prashant Singh}\email{psingh84@ameslab.gov}
\affiliation{Ames Laboratory, U.S. Department of Energy, Iowa State University, Ames, Iowa 50011-3020, USA}
\author{Aftab Alam}\email{aftab@iitb.ac.in}
\affiliation{Department of Physics, Indian Institute of Technology, Bombay, Powai, Mumbai 400076, India} 

\date{\today}

\begin{abstract}
The Jahn-Teller distortion, by its virtue, leads to intriguing electronic properties in transition metal oxides. The charge disproportionation (CD) in Jahn-Teller active systems is a key electronic feature that drives metal-insulator transition (MIT). We demonstrate and quantify it using first-principles calculations combining density functional theory, dynamical mean-field theory, and Monte Carlo simulations. Taking Ca$_2$FeMnO$_6$ as a prototypical example of correlated oxide, our ab-initio study shows that MIT in Ca$_2$FeMoO$_6$ arises from the partial localization of Oxygen ligand holes at alternate Fe sites that controls both charge and magnetic ordering. Interestingly, the band gap was found to be fundamentally controlled by the strength of the charge-transfer energy, and not by the Mott-Hubbard interactions. The novel physics and insights presented in this work reveal promising routes to tune novel electronic functionality in transition metal oxides. 
\end{abstract}

\maketitle


Transition metal oxides (TMO), in recent times, have received renewed attention due to their fascinating electronic properties including metal-insulator transition (MIT) \cite{Varignon2019,Rogge2018,Bennett2022,Balasubramanian2018,Bisogni2016}, colossal magnetoresistance \cite{Jonker1950,Volger1954,Wollan1955} and high-T$_{c}$ superconductivity \cite{Bednorz1986,Wu1980,Schilling1993}. Owing to the presence of strong electron-correlation, the local environment-dependent cooperative phenomenon arising from Jahn-Teller distortion is often found at the heart of most electronic changes \cite{Varignon2019,Varma1988}. The Jahn-Teller active systems also show specific ordering patterns arising from various electronic mechanisms. Notably, the description of MIT in charge disproportionated (CD) mixed valance states has been a long-standing challenge \cite{Imada1998,Barman2000,Mizokawa2000,Johnston2014,Park2012}. 

MIT in high-valence TMOs, distinct from the conventional Mott-insulator, is attributed to the formation of an inhomogeneous charge-disproportionated state controlled by strong Hund’s coupling \cite{Varma1988,Takano1981,Alonso1999,Mizokawa2000}. The CD cations with different charge states, ionic sizes and bond lengths are usually accommodated by Jahn-Teller distortion  \cite{Alonso1999,Mazin2007}. The CaFeO$_3$ and CaCu$_3$Fe$_4$O$_{12}$ are few example where  charge-disproportionated high-valence Fe$^{4+}$ (Fe$^{3+}$ and Fe$^{5+}$) shows three-dimensional rock-salt-type ordering \cite{Woodward2000,Yamada2008,Shimakawa2008}. While stoichiometric Ca$_{2}$FeMnO$_{6}$ (CFMO) is an archetypal example with intriguing two-dimensional checkerboard-type ordering. Multiple transitions including structural (P2$_1$/c to P${\bar 1}$) change followed by low temperature (LT) anti-ferromagnetic charge-disproportionated insulating state makes CFMO even more interesting \cite{Hosaka2015}. Recently, Yang et al.~\cite{Yang2018} performed a detailed electronic-structure study to interpret CFMO. However, the underlying electronic mechanisms that stabilize charge-disproportionated layered ordering largely remain unclear. 

In this letter, we combined first-principles density-functional theory (DFT) method \cite{Kresse1996,Kresse21996,Kresse1993} with dynamical mean-field theory (DMFT)\cite{Blaha2020,Aichhorn2009,Aichhorn2016,Parcollet2015, Werner2006,Seth2016,Held2007,Kraberger2017,Shimakawa2009} and Monte-Carlo simulation to quantify microscopic origin of coexisting unconventional features such as charge-disproportionated MIT phase of CFMO. Our ab-initio calculations show that Jahn-Teller distortion and weak Fe$-$Mn coupling driven charge disproportionation are at the heart of key electronic structure changes displayed in LT CFMO. We also show that the unusual layered ordering of Fe cations facilitates ligand hole-driven CD in CFMO. We also uncover an ordering mechanism where canted AFM ordering is driven by strong competition between super-exchange (SE) and direct-exchange (DE) interaction. 

To understand the structural link to unusual layered ordering, the room temperature (RT) P2$_1$/c (all Fe sites equivalent) and the charge-disproportionated LT P${\bar 1}$ structure observed below 95K were fully relaxed by using PBEsol exchange-correlation functional as implemented within the Vienna Ab-initio Simulation Package (VASP) \cite{Kresse1996,Kresse21996,Kresse1993}. The DFT$+$U method within spin-polarized calculation was done to stabilize Fe cations with LT antiferromagnetic (AFM) order in high-spin states. Calculated structural parameters show fair agreement with experiments (see Fig. S1 and Table S1 of supplementary information (SI) \cite{supp} for more detail). Further computational details are given in SI\cite{supp}; also see Refs.\cite{Perdew1996,Anisimov1997,Dudarev1998}.


\begin{figure*}[t]
\includegraphics[scale=0.31]{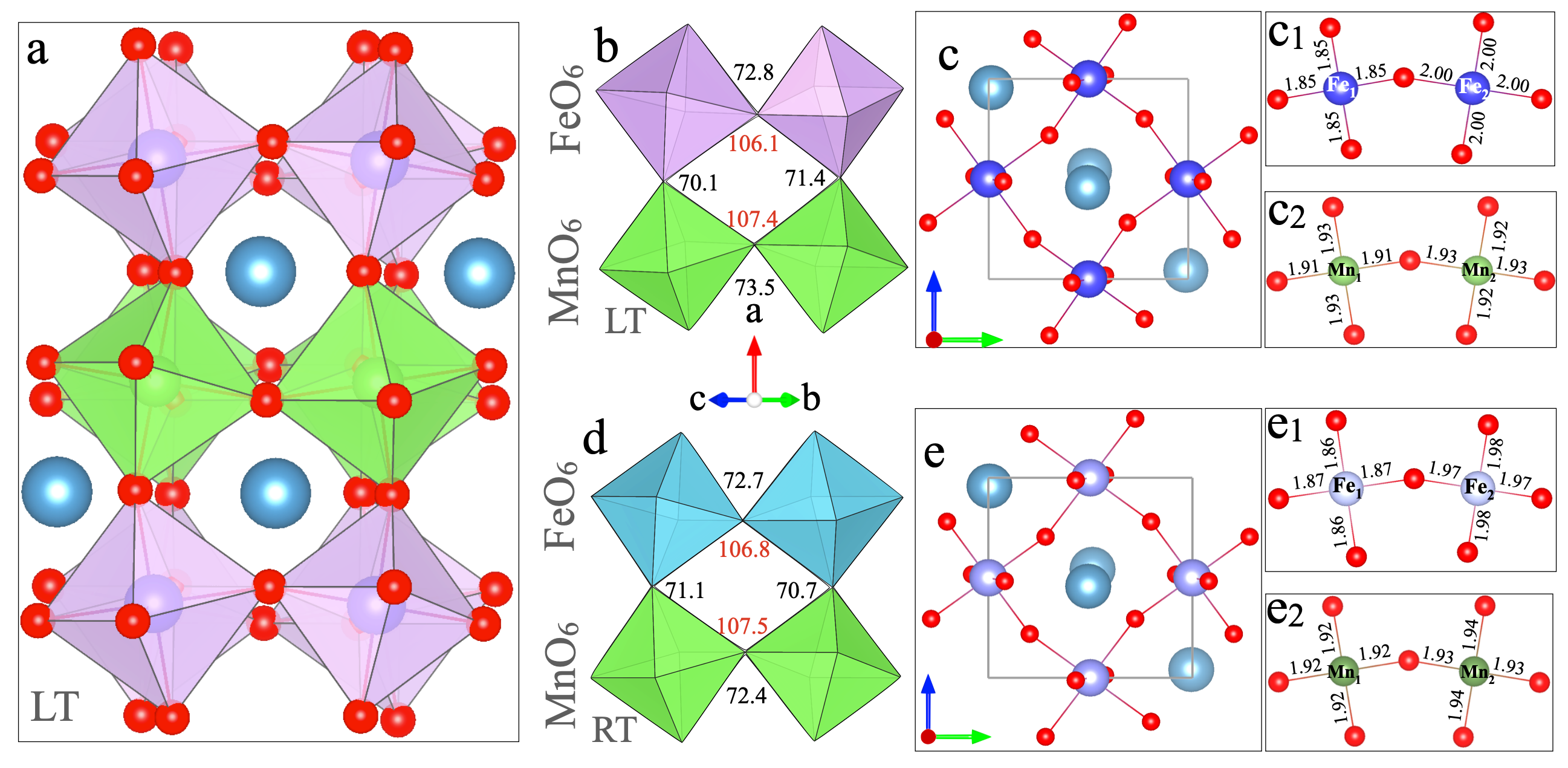}
\caption{(a) Crystal structure (P${\bar 1}$) of Ca$_{2}$FeMnO$_{6}$ at 2 K. Cooperative tilting and bond-lengths arrangement of Fe and Mn octahedra at (b,c, c$_1$, c$_2$) 2 K (P${\bar 1}$) and (d,e, e$_1$, e$_2$) 300 K (P2$_1$/c). All the displayed bond angles/lengths are obtained by theoretically optimizing the corresponding experimental structures at 2 K and 300 K.}
\label{femnd}
\end{figure*}

Figure~\ref{femnd} shows the crystal structure of CFMO including cooperative octahedra tiltings (Fig.~\ref{femnd}(b,d)), (100) projected Fe/Mn-octahedra in their respective phase (Fig.~\ref{femnd}(c,e)), and corresponding Fe/Mn-O bond-lengths (Fig.~\ref{femnd}(c$_1$,c$_2$,e$_1$,e$_2$)) at 2 and 300 K. As shown in Fig.~\ref{femnd}(b,d), tilting mode distortions in both FeO$_6$ and MnO$_6$ octahedra increases as we go from RT to LT phase, which is obvious through bond-angle change. On the contrary, only Fe$-$O bond distances (Fig.~\ref{femnd}(c$_1$, e$_1$)) show visible change in  LT phase while Mn-O bond distance (Fig.~\ref{femnd}(c$_2$, e$_2$)) remain unchanged in both the phases as shown by in-plane projection. For the LT phase, the planar Fe$-$O bond lengths are (2.00; 2.01$~\AA$) whereas the apical bond lengths are 1.95$~\AA$. However, the planer and axial bond length for Fe$_{2}$ octahedra are 1.85$~\AA$ and 1.92$~\AA$ respectively. Clearly, Fe$^{3+}$ (or Fe$_2$) octahedra shows tetragonal compression whereas Fe$^{5+}$ (or Fe$_1$) octahedra shows tetragonal elongation.  In contrast, for Mn$_{2}^{4+}$ octahedra which is connected with Fe$^{3+}$ octahedra, the planner Mn$-$O bond lengths (plane containing Mn ions) are 1.91$~\AA$ and 1.92$~\AA$ whereas axial bond lengths (which connects with Fe octahedra) are 1.93$~\AA$. For Mn$_{2}^{4+}$ (connected with Fe$^{5+}$ octahedra), however, the planner and axial Mn$-$O bond lengths are (1.94$~\AA$, 1.93$~\AA$) and 1.88$~\AA$, respectively.  Bond angles between various TM-O-TM and atom-centered octahedral volumes are detailed in footnote \onlinecite{footnote0}. All the TM-O-TM bond angles varies between 89$^{o}$ to 91$^{o}$. These tilting and breathing mode distortions make LT CFMO a Jahn-Teller active system.



\begin{figure*}[t]
\includegraphics[scale=0.4]{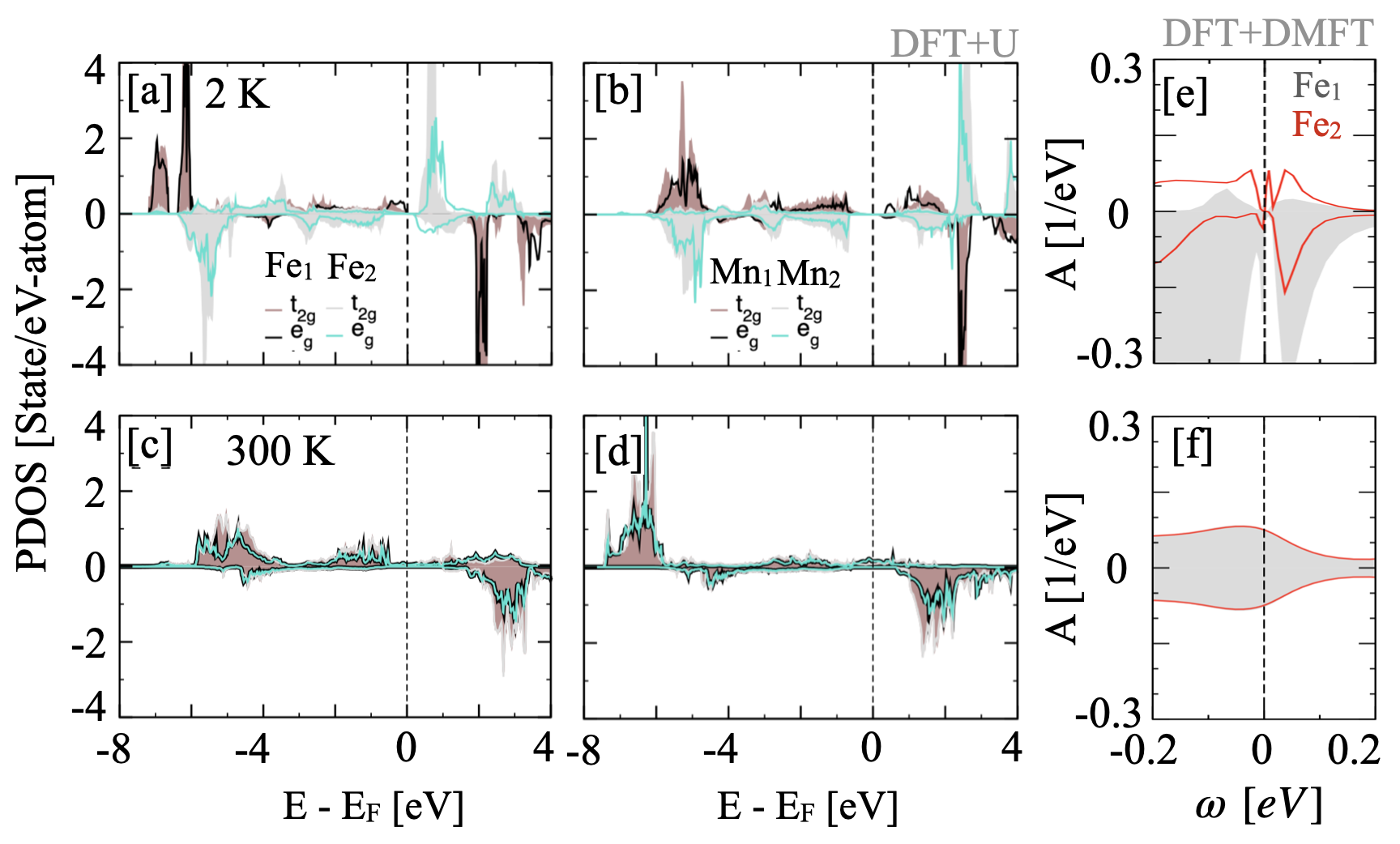}
\caption{For CFMO, the (Fe, Mn) partial density of states for (a,b) LT (2 K), (c,d) RT (300 K) structures. (e,f) DFT+DMFT calculated spectral functions reconfirming insulating (metallic) behavior at LT (RT). See Fig. S11\cite{supp} for full spectral function.}
\label{femnd1}
\end{figure*}

\begin{figure}[b]
\includegraphics[scale=0.35]{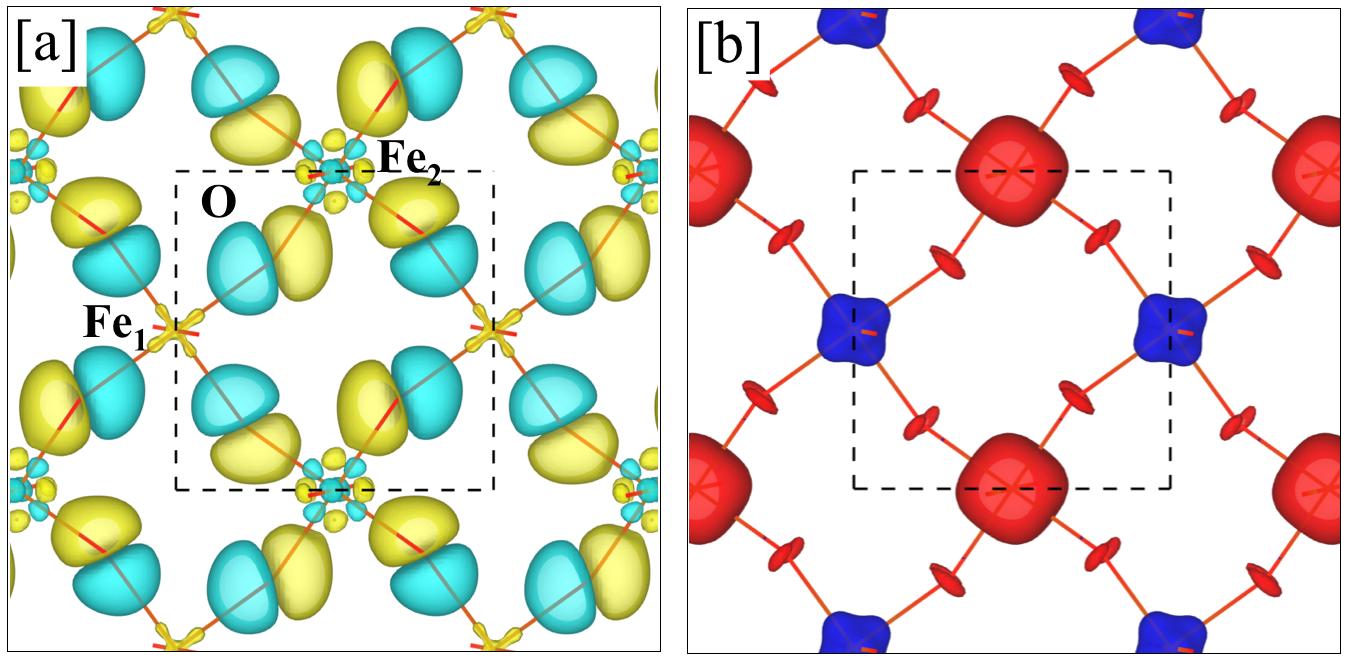}
\caption{(a) Charge (orbital) and (b) magnetic ordering on FeO$_{6}$ octahedral layer for LT structure of CFMO.}
\label{femnd_charge}
\end{figure}

Figure~\ref{femnd1}(a-d) show the Fe- and Mn- partial density of states of CFMO for 2 K and 300 K structures within DFT+U scheme. Application of onsite Hubbard potential ($+$U) splits the $d$-band into an upper and a lower Hubbard band. For Fe$^{3+}$, the lower Hubbard band (Fig. \ref{femnd1}(a)) lies below O$-2p$ level (Fig. \ref{femnd1}(c)) while for Fe$^{5+}$ the Hubbard bands are located at higher energy as compared to that of Fe$^{3+}$. This makes the electron transfer energy from lower Hubbard band to oxygen 2$p$ level negative and the system gains energy through spontaneous transfer of electron from oxygen 2$p$ level to the $d-$level of Fe$^{3+}$ leaving behind a hole($\underline{L}$) on the ligand $p-$orbital (see O$-p$ and Fe$-d$ resolved fat-bands in Fig. S2-S7 \cite{supp} at LT and RT). This spontaneous charge transfer from ligand to the TM$-d$ orbitals gives rise to spontaneous CD at Fe sites (d$^{n}\rightarrow $ d$^{n+1}\underline{L}$), i.e., self-doping [c]. The breathing mode distortion, i.e., tetragonal elongation at Fe$_1$  (Fe$^{5+}$) and compression Fe$_2$ (Fe$^{3+}$), move d$_z^{2}$ orbital at Fe$^{3+}$ site above d$_{x^{2}-y^{2}}$ while it is reversed for Fe$^{5+}$. This cooperative long range Jahn-Teller distortions, d$_z^{2}$ and d$_{x^{2}-y^{2}}$ are expected to be occupied at Fe$^{3+}$ and Fe$^{5+}$ sites, respectively. This alternate occupation of e$_g$ orbitals at two Fe-sites gives rise to AFM-type magnetic order (see Fig.~\ref{femnd_charge}), which gives rise to the charge-disproportionated MIT phase. The electronic occupation for each d-orbital of Fe and Mn in the LT phase of CFMO is shown in SI\cite{supp}. \textcolor{black}{We also simulated the total energy change vs. breathing distortion for FM and AFM phase, as detailed in SI\cite{supp}.} 

The magnitude of the energy gap in the MIT phase is proportional to the charge transfer energy required to move ligand holes to the nearest Oxygen. Because the value of charge transfer energy is less than Hubbard onsite Coulomb potential U, the insulating phase lies in the charge transfer insulating region of the Zaanen-Sawatzky-Allen (ZSA) scheme \cite{Zaanen1985}. This insulating phase is often termed a negative charge transfer insulator. The energy of the lower Hubbard band of Fe$_{2}$ ion and O-$2p$ levels coincide, giving rise to strong hybridization mediated by shorter bond lengths at Fe$^{5+}$ site. With further increase in temperature, the ligand holes gain enough energy to overcome the energy barrier to be mobile, giving rise to a metallic phase as shown by PDOS in Fig.~\ref{femnd1}(e-h).

\textcolor{black}{To further confirm our single particle DFT+U results, we performed DMFT calculations on a Projective Wannier function basis for Fe-$d$ orbitals, which are known to capture dynamic correlation effects. The spectral functions obtained from our DFT+DMFT calculations in both LT and RT are shown in Fig.~\ref{femnd1}(e,f). Details of Wannier projections for DMFT calculations are given in SI\cite{supp}. The spectral function for the RT phase [Fig.~\ref{femnd1}f] shows a metallic character with no charge disproportionation for both impurities Fe$_1$ and Fe$_2$. Impurity charges from DMFT show occupancies of 4.1$e$ on each inequivalent Fe-site. The paramagnetic character of RT phase is confirmed by the symmetric spectral functions in both up and down spin channels (represented by $+$ve and $-$ve sectors of A), while interesting features are observed for LT phase. 
The spectral functions [Fig.~\ref{femnd1}(e)] for LT phase show an insulating behavior with a very small band gap at chemical potential, which is similar to DFT$+$U. However, DMFT shows much better band renormalisation than static DFT+U calculations. Notably, we also observed a distinct CD in the LT state in Fe$_1$ and Fe$_2$ spectral functions, and impurity charges. The impurity charges on Fe$_1$ and Fe$_2$ are 5.9$e$ and 4.1$e$ respectively. The LT structure also shows asymmetry in up and down spin channels confirming an observed AFM  state \cite{Hosaka2015} with magnetic moment on Fe$_1$ oriented opposite to Fe$_2$.} 
\textcolor{black}{The Wannier moments oscillate with DMFT cycles which further affirms AFM behavior \cite{banerjee1, banerjee2} where the magnitude of the Wannier moments are 3.8$\mu_B$ and 3.0$\mu_B$ on Fe$_1$ and Fe$_2$, respectively. Incidentally, the initial Wannier projections from DFT calculations already reflect CD with 4.3$e$ and 3.9$e$ on Fe$_1$ and Fe$_2$ respectively, which gets further enhanced by DMFT renormalisation and corresponding charge transfer from O$-p$ orbitals creating ligand holes.}

Figures \ref{femnd_charge}(a) and \ref{femnd_charge}(b) show the charge and magnetic density at FeO$_{6}$ atomic layers, where Fe$_2$ is more strongly hybridized with oxygen than Fe$_1$ (this is also reflected in change in charge density plots in Fig. S12~\cite{supp}). In the former, orbital on Fe$_1$ site is predominantly d$_{x^2-y^2}$, while the orbital on Fe$_2$ is d$_{z^2}$. The colour represents excess (yellow) and deficiency (turquoise) of electrons for up and down-spin respectively. This leads to strongly localized charges at Fe$_2$ and hence a higher magnetic moment. Figure ~\ref{femnd_charge}(b) indicates a more localized nature of Fe$_1$ orbitals inferring higher on-site coulomb repulsion. This drives our choice of Hubbard U to be 4.5 eV and 2.5 eV for Fe$_1$ and Fe$_2$, respectively. This unique charge and magnetic moment relation makes CFMO more interesting.

It is plausible that alternate occupation of e$_g$ orbital at charge-disproportionated Fe (see Fig. \ref{femnd_charge}) may lead to ferromagnetic (FM) alignment [c-K-K], in order to gain onsite Hund's energy (J$_H$) with exchange parameter $J=-\frac{2t^2}{U-J_H}$ (for Mott insulator) or $J=-\frac{3J_H}{2\Delta_{CT}}\cdot \frac{2t^{2}_d}{\Delta_{CT}+\frac{U_p}{2}}$ (for charge-transfer insulator). However, TM$-$O$-$TM angle at 2 K deviates from perfect 180$^\circ$ and hence the resulting magnetic interaction deviates from that expected for pure 180$^{\circ}$ or 90$^{\circ}$. Thus, the $d-d$ orbital overlaps are not only mediated by $p_\sigma$ orbitals but also through $p_\pi$ orbitals. As a result, half-filled orthogonal $e_{g}$ and $t_{2g}$ orbitals (see Fig.~\ref{femnd1}) have considerable overlap with each other via $p_\pi$ orbitals. The increased AFM exchange strength makes AFM preferable to FM. However, the perfect AFM order is not stable due to the presence of quasi-mobile ligand holes. Instead canted AFM would be more favorable as ligand holes can gain enough kinetic energy through hopping among neighboring sites that would reduce the cost of Hund's energy. This mechanism of magnetic exchange interaction due to mobile or bound carriers is mediated by the double exchange (DE) due to ligand hole causing the charge-disproportionation. The competition between super-exchange and DE interaction gives rise to canted AFM alignment in LT phase, as seen in Table~S2\cite{supp}. At Mn$^{4+}$ sites, the e$_g$ orbitals remain unoccupied and interaction between half-filled $t_{2g}$ orbitals leads to AFM alignment, while SE interaction between half filled $e_{g}$ orbitals of Fe and empty $e_{g}$ orbitals of Mn shows weak FM coupling as shown in Fig.~\ref{femnd1} and \ref{femnd_charge}.

\begin{figure*}
\includegraphics[scale=0.45]{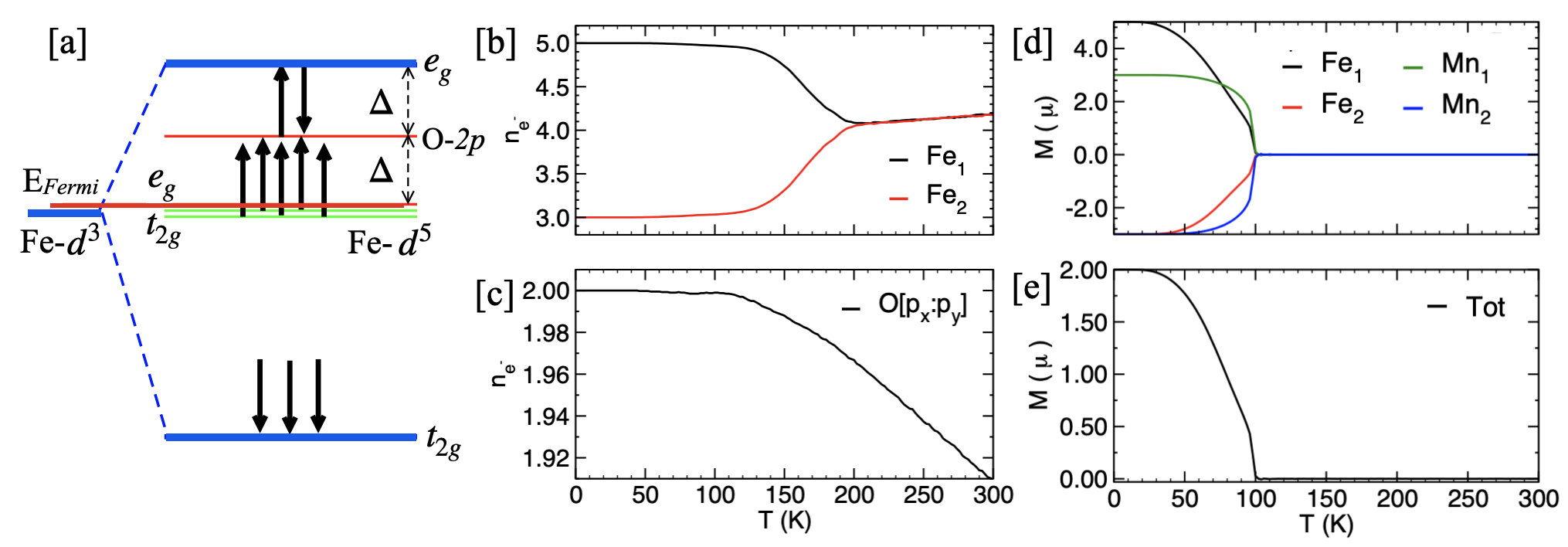}
\caption{For Ca$_2$FeMnO$_6$, (a) schematic of energy level diagram showing charge transfer ($\Delta$) between $d_1$ and $d_2$ cation with $\Delta_1$ =$-\Delta_2$, depicting that energy gain is equal to energy loss at equilibrium. (b, c) Temperature dependence of the number of electrons at Fe$_{1}$, Fe$_{2}$, and O sites. (d, e) T-dependence of  atom-projected and net magnetization (M).}
\label{structures}
\end{figure*}


Among different TM ions in CFMO, only four indirect exchange interactions are possible through the intermediate non-magnetic oxygen atom, namely, Fe$^{3+}$-O-Fe$^{5+}$, Fe$^{3+}$-O-Mn$^{4+}$, Fe$^{5+}$-O-Mn$^{4+}$ and Mn$^{4+}$-O-Mn$^{4+}$ (see Table S3 of SM~[\cite{supp}] for calculated exchange parameters). It is evident from Fig.~\ref{femnd} that the Fe$-$O$-$Fe and Mn$-$O$-$Mn lie on crystallographic (011)-plane whereas the Fe$-$O$-$Mn bonds are along (100).  Because of strong onsite coulomb repulsion, hopping to the Fe$_1$ electronic level is unlikely from neighboring atoms. While the coulomb interaction on Fe$_2$ atom is weaker as it has almost empty $d_{yz}^{\uparrow}$. Therefore, the hopping from p$_y$ or p$_z$ level of oxygen can give FM coupling with Fe$_1$. The half-filling of $d_{x^2-y^2}$ orbitals at Mn$_2$, similar to Fe$_1$, creates an SE mediated AFM ordering as depicted by Goodenough and Kanamori \cite{Mizokawa2000,Bersuker2006,Heitler1927}. On the other hand, the d$_{x^2-y^2}$ orbitals at Fe$_2$ are fully occupied while e$_{g}$ in Mn$_1$ show partial filling that results in a very weak FM interaction. However, half-filled t$_{2g}$ orbitals in Mn leads to a stronger AFM interaction. In general, the SE interactions are stronger when TM$-$O$-$TM bond angles are 180$^\circ$. But because of the deviation in the present case, different types of interaction arise which creates a non-collinear magnetic structure (corresponding moments at each TM ion are given in Table S4 of SM\cite{supp}). 

In CFMO, CaFeO$_3$, and other related oxides, it has been observed that Fe$^{4+}$ transfers a $d-$electron to its neighboring Fe atoms \cite{Woodward2000}, which creates an alternating checkerboard pattern of  Fe$^{3+}$ and Fe$^{5+}$ pairs in the 2D FeO$_{6}$ octahedral layers. This mechanism of CD at FeO$_6$ octahedra is closely related to the crystal field stabilization energy (CFSE). Ideally, for $d^3$, $d^{4}$ and $d^{5}$ in octahedral environment, CFSE (i.e., $\Delta_{o}$) is $-\frac{6}{5}\Delta_o$, $-\frac{3}{5}\Delta_o$ and 0, respectively. Notably, CFSE of two Fe$^{4+}$ is the sum of Fe$^{3+}$ and Fe$^{5+}$, which means that energetic preference for charge splitting in CFMO has no energy cost at LT. On the other hand, the MnO$_6$ octahedral layers are ordered antiferomagnetically like those in CaMnO$_3$, where Mn exists in $d^{3+}$ state. This would require $\frac{1}{5}\Delta_o$ more energy for CD between $d^2$ and $d^4$. This higher energy cost provides Mn with a more robust oxidation state than Fe in CMFO, leading to no CD for Mn in LT-phase. This suggests that the weak coupling between Fe-Mn layers and high effective ligand hole spin interactions CD at Fe cations in LT phase is the key reason for the appearance of 2D checkerboard type AFM order in CFMO. 

Next, we constructed an extended Hubbard model for CFMO (Eq. \ref{TBH} below) and combined it with  Monte Carlo (MC) simulation to understand the role of CD on magnetization and charge ordering behavior. 


\begin{eqnarray} 
H&=&-\sum_{i,j}t_{ij}^{pd}(c_{i\sigma}^{\dagger p}c_{j\sigma}^{d}+h.c)+\sum_{i,\alpha}U^{d}_i\left(n_{i\alpha\uparrow}^{d}-\frac{1}{2}\right)\times\nonumber\\
&&\left(n_{i\alpha\downarrow}^{d}-\frac{1}{2}\right)-\sum_{i\sigma}\mu^{d}_i n^d_{i\sigma}-\mu_p\sum_{i\sigma}n^p_{i\sigma}
\label{TBH}
\end{eqnarray} 
where, $\mu$ is the chemical potential of ($p,d$) orbitals. This Hamiltonian describes 5-fold degenerate $d$-orbitals at each Fe site connected via a single $p$-orbital. $i$ and $j$ indicate different sites while $\alpha$ denotes various $d$-orbitals. Maximum occupancy of $d$-orbitals can be $1$, implying half-filled. Let $n_1$, $n_2$ and $n_p$ are the number of electrons at Fe$_1$, Fe$_2$ and intermediate O-p orbitals, with potential energy expressed as $ V=\left( \frac{5}{2}-n_1\right)\frac{U_1}{2}+\left( \frac{5}{2}-n_2\right)\frac{U_2}{2} -n_1\mu_1-n_2\mu_2-n_p\mu_p $. This relation is valid when $n_1$ and $n_2$ are less than or equal to 5, implying $d$-orbitals to be half-filled or less than half-filled. If $d$-orbitals become more than half-filled, the Hubbard  terms change to  $\left(n_1-\frac{15}{2}\right)\frac{U_1}{2}$ and $\left(n_2-\frac{15}{2}\right)\frac{U_2}{2}$. Both Fe$_1$ and Fe$_2$ along with Mn are found to couple antiferromagnetically \cite{Hosaka2015}.

{\par} \textcolor{black}{ We chose collinear spin configuration as magnetic ground state and the same Hubbard parameters as obtained from DFT calculations to construct Eq.~(\ref{TBH}). The chemical potential ($\mu$) for Fe$_1$, Fe$_2$ and oxygen was set to 0.5, 1.5 and 3.0 eV, respectively. While the choice of crystal-field values ($\Delta$) at Fe$_1$ (0 eV) and Fe$_2$ (1.2 eV) is driven by the fact that Fe$_1$ does not show any crystal-field splitting in DFT while Fe$_2$ had large splitting. The hopping parameters ($t_{ij}$) were fixed at 3.2 eV. The MC simulation was performed over an 8$\times$8$\times$8 supercell (a total of 4096 atoms) where Ca-atoms were ignored as they do not play any active role in charge-transfer or magnetic transition (further details about MC simulation are provided in the footnote~[\onlinecite{footnote1}]). Based on DFT calculated $t_{ij}$, U, $\mu$, and $\Delta$ values, we show a representative schematic of the energy level diagram in Fig.~\ref{structures}(a) for CFMO in LT Phase. These details were used within the MC scheme to simulate the electron dynamics with temperature.}


\textcolor{black}{ The large crystal-field splitting between t$_{2g}$ and e$_g$ states for Fe$_2$ (+5) prohibits the electron hopping between these states. 
However, the presence of O-2$p$ states at/near the Fermi-level (see Fig. S4, S5, S6 of SI\cite{supp}) between Fe$_1-$e$_g$ and Fe$_2-$t$_{2g}$ provides energetically favorable exchange path for electrons in Fe$_1-d-$e$_g$ states to hop, i.e., Fe$_1-d$-O$-p-$Fe$_2-d$. 
Further, our MC simulations show that the charge ordering on Fe$_1$ ($+$3) and Fe$_2$ ($+$5) sites disappears at/around 200 K, where both Fe-atoms approximately reach $+4$ oxidation state (see Fig.~\ref{structures}(b)), which agrees fairly well with experiments \cite{Hosaka2015}. For LT (below 150K), the in-plane Oxygen connected to Fe loses some charge creating hole states (see Fig.~\ref{structures}(c)), which play a crucial role in driving CMFO towards a metallic phase at LT, resulting in a MIT. This is in concurrence with our DFT and DFMT simulation (see Fig.~2). 
Finally, the T-dependence of atom-projected and net magnetization in Fig.~\ref{structures}(d,e) also shows a clear transition at $\sim$ 100 K, as observed experimentally \cite{Hosaka2015}.}

In summary, we offer deep microscopic insights into the charge- and magnetic-ordering mechanisms in Ca$_2$FeMnO$_6$. The CD in the LT phase plays a critical role in mediating such ordering arising from Jahn-Teller distortion, which is induced by the partial localization of O$-2p$ ligand holes at alternate Fe sites. This eventually leads to an unusual charge transfer band gap. The temperature-dependent Monte-Carlo simulations further show the correlation between the charge transfer gap and MIT. We show that the inter-site interaction, responsible for charge transfer between Fe$_1$ and Fe$_2$, is much more than the Fe$-$O interaction. Our simulation suggests that self-hole doping is highly likely to trigger such a transition in CFMO. Interestingly, this is yet another unique system in which the amount of ligand-hole is a key factor controlling MIT, where band-gap is fundamentally controlled by the strength of the charge-transfer energy close to the MIT, and not by the Mott-Hubbard interactions as originally thought for Mottronics applications. Such charge disproportionation and MIT can, in principle, originate via composition, pressure and/or stoichiometry modulations in other perovskite classes as well.

{\it Acknowledgement:~} A.A. acknowledges DST-SERB (Grant No. CRG/2019/002050) for funding to support this research. HB acknowledges computing support from Odyssey and Sulis HPC clusters. This work was supported by the U.S. Department of Energy (DOE), Office of Science, Basic Energy Sciences, Materials Science and Engineering Division. The research is performed at the Ames Laboratory, which is operated for the U.S. DOE by Iowa State University under contract DE-AC02-07CH11358.


\begin{thebibliography}{56}  
\bibitem{Imada1998}M. Imada, A. Fujimori, and Y. Tokura, Rev. Mod. Phys. {\bf70}, 1039 (1998). 
\bibitem{Barman2000} S. R. Barman, A. Chainani, and D. D. Sarma, Phys. Rev. B {\bf 49}, 8475 (1994). 

\bibitem{Johnston2014} S Johnston, A. Mukherjee, I. Elfimov, M. Berciu, and G.A. Sawatzky, Charge Disproportionation without Charge Transfer in the Rare-Earth-Element Nickelates as a Possible Mechanism for the Metal-Insulator Transition, Phys. Rev. Lett. {\bf 112}, 106404 (2014).
\bibitem{Park2012} H. Park, A. J. Millis, and C. A. Marianetti, Phys. Rev. Lett. {\bf 109}, 156402 (2012). 

\bibitem{Varignon2019} J. Varignon, M. Bibes,  and A. Zunger, Origin of band gaps in 3d perovskite oxides. Nat Commun {\bf 10}, 1658 (2019).
\bibitem{Dalpian2018} G. M. Dalpian, Q. Liu, J. Varignon, M. Bibes, and A. Zunger, Bond disproportionation, charge self-regulation, and ligand holes in $s-p$ and $d-$electron ABX$_{3}$ perovskites by density functional theory, Phys. Rev. B {\bf 98}, 075135 (2018).

\bibitem{Rogge2018} P. C. Rogge et al. Electronic structure of negative charge transfer CaFeO$_{3}$ across the metal-insulator transition, Phys. Rev. Mater {\bf 2}, 015002 (2018).
\bibitem{Bennett2022} M.C. Bennett, G. Hu, G. Wang, O. Heinonen, P.R.C. Kent, J. T. Krogel, and P. Ganesh, Origin of metal-insulator transitions in correlated perovskite metals, Phys. Rev. Research {\bf 4}, L022005 (2022).
\bibitem{Balasubramanian2018} P. Balasubramanian, Electronic structure of Pr$_{2}$MnNiO$_{6}$ from x-ray photoemission, absorption and density functional theory, J. Phys.: Condens. Matter {\bf 30} 435603 (2018).
\bibitem{Bisogni2016} V. Bisogni, S. Catalano, R. Green et al. Ground-state oxygen holes and the metal-insulator transition in the negative charge-transfer rare-earth nickelates, Nat Commun {\bf 7}, 13017 (2016).
\bibitem{Jonker1950} G.H. Jonker, and J.H. Van Santen, Ferromagnetic compounds of manganese with perovskite structure, Physica {\bf 16} (3), 337 (1950).
\bibitem{Volger1954} J. Volger, Further experimental investigations on some ferromagnetic oxidic compounds of manganese with perovskite structure, Physica {\bf 20} (1), 49-66 (1954).
\bibitem{Wollan1955} E.O. Wollan, and W.C. Koehler, Neutron Diffraction Study of the Magnetic Properties of the Series of Perovskite-Type Compounds La$_{(1-x)}$Ca$_x$MnO$_{3}$"; Physical Review {\bf 100} (2), 545 (1955)
\bibitem{Bednorz1986} J. G. Bednorz, and K. A. Mueller, Possible high T$_{C}$ superconductivity in the Ba-La-Cu-O system, Z. Phys. B. {\bf 64} (2). 189-193 (1986).
\bibitem{Wu1980} M. K. Wu et al. Superconductivity at 93 K in a New Mixed-Phase Y-Ba-Cu-O Compound System at Ambient Pressure, Ten Years of Superconductivity: 1980-1990, Perspectives in Condensed Matter Physics, Dordrecht: Springer Netherlands, 7, pp. 281-283 (1993).
\bibitem{Schilling1993} A. Schilling, M. Cantoni, J.D. Guo, and H.R. Ott, Superconductivity above 130 K in the Hg-Ba-Ca-Cu-O system, Nature {\bf 363} (6424), 56-58 (1993).
\bibitem{Hosaka2015} Y. Hosaka, N. Ichikawa, T. Saito, P. Manuel, D. Khalyavin, J. P. Attfield, and Y. Shimakawa, Two-Dimensional Charge Disproportionation of the Unusual High Valence State Fe$^{4+}$ in a Layered Double Perovskite, J. Am. Chem. Soc. {\bf 137} (23), 7468-7473 (2015). 


\bibitem{Yang2018} K. Yang, D.I. Khomskii, and H. Wu, Unusual layered order and charge disproportionation in the double-perovskite compound  Ca$_2$FeMnO$_6$, Phys. Rev. B {\bf 98}, 085105 (2018).

\bibitem{Mizokawa1991} T. Mizokawa, H. Namatame, A. Fujimori, K. Akeyama, H. Kondoh, H. Kuroda, and N. Kosugi, Origin of the band gap in the negative charge-transfer-energy compound NaCuO$_2$, Phys. Rev. Lett. {\bf 67}, 1638 (1991)

 \bibitem{Hubbard1963} J. Hubbard, Electron Correlations in Narrow Energy Bands Proceedings of the Royal Society of London, {\bf 276} (1365), 238257 (1963).
\bibitem{Bocquet1992} A. E. Bocquet, T. Mizokawa, T. Saitoh, H. Namatame, and A. Fujimori, Electronic structure of 3d-transition-metal compounds by analysis of the 2p core-level photoemission spectra, Phys. Rev. B {\bf 45}, 3771 (1992).
\bibitem{Zaanen1985} J. Zaanen, G.A. Sawatzky, and J.W. Allen, Band gaps and electronic structure of transition-metal compounds, Phys. Rev. Lett. {\bf 55}, 418 (1985).



\bibitem{Varma1988} C.M. Varma, Missing valence states, diamagnetic insulators, and superconductors, Phys. Rev. Lett. {\bf 61}, 2713 (1988).
\bibitem{Takano1981} M. Takano, J. Kawachi, N. Nakanish, and Y. Takeda, Valence state of the Fe ions in Sr$_{1-y}$La$_{y}$FeO$_{3}$, J. Solid State Chem. {\bf 39}, 75 (1981).
\bibitem{Alonso1999} J. A. Alonso, J. L. Garcia-Munoz, M. T. Fernandez-Díaz, M. A. G. Aranda, M. J. Martinez-Lope, and M. T. Casais, Charge Disproportionation in RNiO$_{3}$ Perovskites: Simultaneous Metal-Insulator and Structural Transition in YNiO$_{3}$, Phys. Rev. Lett. {\bf 82}, 3871 (1999).

\bibitem{Mazin2007} I. I. Mazin, D. I. Khomskii, R. Lengsdorf, J. A. Alonso, W. G. Marshall, R. M. Ibberson, A. Podlesnyak, M. J. Martínez-Lope, and M. M. Abd-Elmeguid, Phys. Rev. Lett. {\bf 98}, 176406 (2007). 

\bibitem{Yamada2008} I. Yamada, K. Takata, N. Hayashi, S. Shinohara, M. Azuma, S. Mori, S. Muranaka, Y. Shimakawa, M. Takano, A perovskite containing quadrivalent iron as a charge-disproportionated ferrimagnet, Angew. Chem. – Int. Ed. {\bf 47}, pp. 7032-7035 (2008).
\bibitem{Shimakawa2008} Y. Shimakawa, A-site-ordered perovskites with intriguing physical properties, Inorg. Chem. {\bf 47}, 8562-8570 (2008).

\bibitem{Mizokawa2000} T. Mizokawa, D.I. Khomskii, and G.A. Sawatzky, Spin and charge ordering in self-doped Mott insulators, Phys. Rev. B {\bf 61}, 11263 (2000).
\bibitem{Bersuker2006} I.B. Bersuker, The Jahn-Teller Effect. Cambridge: Cambridge University Press (2006).
\bibitem{Heitler1927} W. Heitler, and F. London. Wechselwirkung neutraler Atome und homopolare Bindung nach der Quanten-mechanik, Z. Phys. {\bf 44}, 455 (1927).
\bibitem{Kramers1934} H.A. Kramers. L'interaction Entre les Atomes Magntognesdans un Cristal Paramagntique, Physica 1, 182 (1934).
\bibitem{Anderson1950} P.W. Anderson, Antiferromagnetism. Theory of Superexchange Interaction, Phys. Rev. {\bf 79}, 350 (1950).
\bibitem{Anderson1959} P.W. Anderson, New Approach to the Theory of Superexchange Interactions, Phys. Rev. {\bf 115}, 2 (1959).
\bibitem{Ruderman1954} M.A. Ruderman, and C. Kittel, Indirect Exchange Coupling of Nuclear Magnetic Moments by Conduction Electrons, Physical Review {\bf 96} (1), 99-102 (1954).
\bibitem{Kasuya1956} T. Kasuya, Theory of Metallic Ferro- and Antiferromagnetism on Zener's Model, Prog. Theor. Phys. {\bf 16} (1), 45-57 (1956).
\bibitem{Yosida1957} K. Yosida, Interaction between the d -Shells in the Transition Metals. II. Ferromagnetic Compounds of Manganese with Perovskite Structure, Phys. Rev. {\bf 106} (5), 893-898  (1957).
\bibitem{Zener1951} C. Zener, Interaction between the d -Shells in the Transition Metals. II. Ferromagnetic Compounds of Manganese with Perovskite Structure, Phys. Rev. {\bf 82}, 403 (1951).
\bibitem{Anderson1955} P. W. Anderson and H. Hasegawa, Considerations on Double Exchange, Phys. Rev. {\bf 100}, 675, (1955).
\bibitem{Gennes1960} P. G. de Gennes, Effects of Double Exchange in Magnetic Crystals, Phys. Rev. {\bf 118}, 141 (1960).
\bibitem{Kresse1996} G. Kresse, and J. Furthmuller, Efficient iterative schemes for ab initio total-energy calculations using a plane-wave basis set, Phys. Rev. B {\bf 54},11169 (1996).
\bibitem{Kresse21996} G. Kresse, and J. Furthmller, Efficiency of ab-initio total energy calculations for metals and semiconductors using a plane-wave basis set, Comp. Mater. Sci. {\bf 6}, 15 (1996).
\bibitem{Kresse1993} G. Kresse, and J. Hafner, Ab initio molecular dynamics for liquid metals, Phys. Rev. B {\bf 47}, 558 (1993).
\bibitem{supp} See supplemental material at [URL to be added] for auxiliary informations about further computational details, extended Hubbard model,  theoretically optimised structural parameters, magnetic structure and exchange interactions, crystal structures at 300 and 2 K, and different orbital projected band structures.
\bibitem{Kresse1999} G. Kresse and D. Joubert, From ultrasoft pseudopotentials to the projector augmented-wave method, Phys. Rev. B {\bf 59}, 1758 (1999).
\bibitem{Perdew1996} J. P. Perdew, K. Burke, and M. Ernzerhof, Generalized Gradient Approximation Made Simple, Phys. Rev. Lett. {\bf 77}, 3865 (1996).
\bibitem{Anisimov1997} V.I. Anisimov, F. Aryasetiawan, and A.I. Liechtenstein,  First-principles calculations of the electronic structure and spectra of strongly correlated systems: the LDA$+$U method, J. Phys.: Condens. Matter {\bf 9}, 767 (1997).
\bibitem{Dudarev1998} S.L. Dudarev, G.A. Botton, S.Y. Savrasov, C.J. Humphreys, and A.P. Sutton, Electron-energy-loss spectra and the structural stability of nickel oxide: An LSDA$+$U study, Phys. Rev. B {\bf 57}, 1505 (1998).

\bibitem{Blaha2020} P. Blaha, K. Schwarz, F. Tran, R. Laskowski, G. K. H. Madsen, and L. D. Marks, The Journal of Chemical Physics {\bf 152}, 074101 (2020). 
\bibitem{Aichhorn2009} M. Aichhorn, L. Pourovskii, V. Vildosola, M. Ferrero, O. Parcollet, T. Miyake, A. Georges, and S. Biermann, Phys. Rev. B {\bf 80}, 085101 (2009). 
\bibitem{Aichhorn2011} M. Aichhorn, L. Pourovskii, and A. Georges, Phys. Rev. B {\bf 84}, 054529 (2011). 
\bibitem{Aichhorn2016} M. Aichhorn, L. Pourovskii, P. Seth, V. Vildosola, M. Zingl, O. E. Peil, X. Deng, J. Mravlje, G. J. Kraberger, C. Martins, M. Ferrero, and O. Parcollet, Comput. Phys. Commun. {\bf 204}, 200 (2016). 
\bibitem{Parcollet2015} O. Parcollet, M. Ferrero, T. Ayral, H. Hafermann, I. Krivenko, L. Messio, and P. Seth, Comput. Phys. Commun. {\bf 196}, 398 (2015). 
\bibitem{Werner2006} P. Werner and A. J. Millis, Phys. Rev. B {\bf 74}, 155107 (2006). 
\bibitem{Seth2016} P. Seth, I. Krivenko, M. Ferrero, and O. Parcollet, Comput. Phys. Commun. {\bf 200}, 274 (2016). 
\bibitem{Held2007} K. Held, Advances in Physics {\bf 56}, 829 (2007). 
\bibitem{Kraberger2017} G. J. Kraberger, R. Triebl, M. Zingl, and M. Aichhorn, Phys. Rev. B {\bf 96}, 155128 (2017). 

\bibitem{Shimakawa2009} Y. Shimakawa, and M. Takano, Charge Disproportionation and Charge Transfer in A-site Ordered Perovskites Containing Iron, Z. Anorg. Allg. Chem. {\bf 635}, 1882 (2009).

\bibitem{footnote0} Fe$-$O$-$Fe bond angles in Fe-plane are around 153$^{o}$, while Mn$-$O$-$Mn bond angles are about 154$^{o}$. Fe$^{3+}-$O$-$Mn$^{4+}$ bond angle is 156 and Fe$^{5+}-$O$-$Mn$^{4+}$ bond angle is 153 degrees as shown in Fig. (1). Octahedral volumes of Fe$^{3+}$, Fe$^{5+}$ are 10.44 \AA$^{3}$~ and 8.75 \AA$^{3}$, respectively. The Fe$^{5+}$ connected Mn$^{4+}$ octahedral volume is 9.48 \AA$^{3}$, whereas Fe$^{3+}$ connected Mn$^{4+}$ octahedral volume is 9.35 \AA$^{3}$. 

\bibitem{banerjee1} H. Banerjee, H. Schnait, M. Aichhorn, and T. Saha-Dasgupta, Effect of geometry on magnetism of Hund's metals: Case study of  
BaRuO$_3$, Phys. Rev. B 105, 235106 (2022)

\bibitem{banerjee2} H. Banerjee, C. P. Grey, and A. J. Morris, Importance of electronic correlations in exploring the exotic phase diagram of layered Li$_x$MnO$_2$, Phys. Rev. B 108, 165124 (2023)

\bibitem{Goodenough1955} J.B. Goodenough, Theory of the Role of Covalence in the Perovskite-Type Manganites, Phys. Rev. {\bf 100}, 564 (1955).
\bibitem{Goodenough1958} J. B. Goodenough, An interpretation of the magnetic properties of the perovskite-type mixed crystals La$_{1-x}$Sr$_{x}$CoO$_{3-\lambda}$, J. Phys. Chem. Solids {\bf 6} (2-3), 287 (1958).
\bibitem{Kanamori1959} J. Kanamori, Superexchange interaction and symmetry properties of electron orbitals, J. Phys. Chem. Solid {\bf 10}, 87 (1959).
\bibitem{footnote1} In the Monte Carlo simulation, each electrons were treated as spin $1/2$ particle, mimicking an Ising like problem with no charge transfer. Metropolis algorithm was used to identify spin orientation and position of of electrons at each T. 2000 Monte Carlo steps were used for equilibration at each T. U values were chosen to be the same as in our DFT calculations. The chemical potential of oxygen ($\mu_p$) was chosen in such a way that the energy needed for electron transfer ($\Delta$) is relatively small. Exchange interactions, as calculated from DFT, are $J_{Fe_{1}-Fe_{2}}$=-1.45, $J_{Fe_{1}-Mn_{2}}$=-0.84, $J_{Mn_{1}-Fe_{2}}$=0.30 and $J_{Mn_{1}-Mn_{2}}$=-1.86 meV. 
\bibitem{Woodward2000} P.M. Woodward, D.E. Cox, E. Moshopoulou, A.W. Sleight, and S. Morimoto, Structural studies of charge disproportionation and magnetic order in CaFeO$_{3}$, Phys. Rev. B {\bf 62}, 844 (2000).

\end{thebibliography}
\end{document}